\documentclass{PoS}

\usepackage{amsfonts}
\usepackage{amsmath}
\usepackage{amssymb}
\usepackage{mathdots}
\usepackage{graphicx}
\usepackage[algosection,boxruled]{algorithm2e}
\newcommand{\mathR}{\mathbb{R}}
\newcommand{\mathC}{\mathbb{C}}
\newcommand{\eqnref}[1]{(\ref{#1})}
\newcommand{\sign}{{\rm sign}}
\newcommand{\spec}{{\rm spec}}

\graphicspath{{./}{pics/}}

\newtheorem{theorem}{Theorem}
\newtheorem{corollary}{Corollary}

\DontPrintSemicolon
\newcommand{\rr}{\mbox{\rm r}}

\title{Error Bounds for the Sign Function}

\ShortTitle{Error Bounds for the Sign Function}

\author{\speaker{Andreas Frommer} 
         \thanks{This work was supported by Deutsche Forschungsgemeinschaft through the Collaborative Research Centre SFB-TR 55 "Hadron Physics from Lattice QCD"}\\
        Fachbereich C, Mathematik und Naturwissenschaften, Bergische Universit\"at Wuppertal, D-42097 Wuppertal, Germany\\
        E-mail: \email{frommer@math.uni-wuppertal.de}}
\author{Karsten Kahl\\
       Fachbereich C, Mathematik und Naturwissenschaften, Bergische Universit\"at Wuppertal, D-42097 Wuppertal, Germany\\
       E-mail: \email{kahl@math.uni-wuppertal.de}}
\author{Thomas Lippert\\
         J\"ulich Supercomputing Centre, Forschungszentrum J\"ulich GmbH, D-52425 J\"ulich, Germany \\
         E-mail: \email{th.lippert@fz-juelich.de}}
\author{H.~Rittich\\
       Fachbereich C, Mathematik und Naturwissenschaften, Bergische Universit\"at Wuppertal, D-42097 Wuppertal, Germany\\
       E-mail: \email{rittich@math.uni-wuppertal.de}}

\abstract{The Overlap operator fulfills the Ginsparg-Wilson relation exactly and therefore represents an optimal discretization of the QCD Dirac operator with respect to chiral symmetry. When computing propagators or in HMC
simulations, where one has to invert the overlap operator using some iterative solver, one has to approxomate 
the action of the sign function of the (symmetrized)
Wilson fermion matrix $Q$ on a vector $b$ in each iteration. This is usually done iteratively using a `primary' Lanczos iteration. In this process, it is very important to have good stopping criteria which allow to reliably assess the quality of the approximation to the action of the sign function computed so far.     
In this work we show how to cheaply recover a secondary Lanczos process, starting at an arbitrary Lanczos vector of the primary process and how to use this secondary process to efficiently obtain computable error estimates and error bounds
for the Lanczos approximations to $\sign(Q)b$, where the sign function is approximated by the Zolotarev rational approximation.}

\FullConference{XXIX International Symposium on Lattice Field Theory \\
		 July 10-16, 2011\\
		 Squaw Valley, Lake Tahoe, California}

\begin{document}

\section{Introduction}
Overlap fermions as a lattice formulation of QCD respecting chiral symmetry
have been proposed in \cite{Narayanan:2000qx} and been investigated since by many authors.
The overlap operator still represents the discrete Dirac operator which most neatly deals with 
chiral symmetry, fulfilling the Ginsparg-Wilson relation on the lattice exactly.
If $D_W$ describes the hopping part of the standard Wilson fermion matrix and $\kappa_c$ its
critical hopping parameter, the overlap operator is given as
\[
D_O = I + \rho \gamma_5 \sign(Q) \mbox{ with } Q = \gamma_5(I - \frac{4\kappa_c}{3} D_W).
\] 
Herein, $\rho$ is a mass parameter which is close to 1.

A direct computation of $\sign(Q)$ is not feasible, since $Q$ is large and
sparse, whereas $\sign(Q)$ would be full. Therefore,
numerical algorithms which invert systems with the matrix $D_O$ have to follow  an
inner-outer paradigm: One performs an outer Krylov subspace
method where each iteration requires the computation of a matrix-vector
product involving $\sign(Q)$. Each such product is computed through
another, inner iteration using matrix-vector multiplications with $Q$.
In this context, it is very important to be able to assess the accuracy of the computed 
approximation to $\sign(Q)b$ from the inner method, since one can steer the outer method so
as to require less and less accurate computations of $\sign(Q)b$, resulting in substantial 
savings in computational work, see \cite{Cu05a}.

In this work we precisely consider the task of obtaining reliable error estimates and bounds 
when computing approximations for $\sign(Q)b$. Most preferably, we would like to have a precise
{\em  upper bound}, so that a stopping criterion based on that upper bound will guarantee that the
exact error is below this bound. Actually, we will consider the case where the sign function 
$\sign(t)$ is approximated by a rational function $g(t)$, the Zolotarev approximation. This approach 
has established itself as the method of choice, since the multishift cg method allows
for an efficient update of the iterates, involving only short recurrencies and thus few memory \cite{vdE02}. 

Usually, one fixes the rational Zolotarev approximation $g(Q)b$ such that the error w.r.t. the sign function is less than $\epsilon_1$ on the spectrum of $Q$.  An error bound $\epsilon_2$ for the approximation of $g(Q)b$ then results in an overall error bound $\epsilon_1 + \epsilon_2$ w.r.t.\ $\sign(Q)b$.

\section{Lanczos process and Lanczos approximations} \label{lanczos:sec}

Assuming that $v_1 \in \mathC^{n}$ is normalized to $\|v_1\|_2 = 1$, 
the Lanczos process computes orthonormal vectors $v_1,v_2,\ldots$ such that $v_1,\ldots,v_m$ form an orthonormal basis of the nested sequence of Krylov subspaces $K_m(Q,v_1)$, $m=1,2,\ldots$. It is given here as Algorithm~\ref{algo:lanczos}. 

\begin{algorithm}
  \caption{Lanczos process with maztrix $A$ and starting vector $v_1$}
  \label{algo:lanczos}
  choose $v_1$ such that $\| v_1 \| = 1 $ \;
  let $\beta_0 := 0$, $v_0 := 0$ \;
  \For{ $j = 1, \dots, m$ }{
    $w_j = A v_j - \beta_{j-1} v_{j-1}$ \;
    $\alpha_j = v_j^*w_j$ \; 
    $w_j = w_j - \alpha_j v_j$ \;
    $\beta_j = \| w_j \|_2$ \;
    \lIf{ $\beta_j = 0$ }{
      stop
    } \;
    $v_{j+1} = (1/ \beta_j) \cdot w_j $
  }
\end{algorithm}

The Lanczos process can be summarized via the {\em Lanczos relation}
\begin{equation} \label{lanczos_rel:eq}
A V_m = V_{m+1}T_{m+1,m} = V_m T_m + \beta_{m}\cdot e_m^*v_{m+1},
\end{equation}   
where $V_m = [v_1|\ldots|v_m] \in \mathbb{C}^{n \times m}$ is the matrix containing the Lanczos vectors,
$e_m = (0,\ldots,0,1)^* \in \mathC^m$ 
and
\[
  T_{m+1,m} = \begin{bmatrix}
    \alpha_1 & \beta_1 \\
    \beta_1 & \alpha_2 & \ddots \\
    & \ddots & \ddots & \beta_{m-1} \\
    & & \beta_{m-1} & \alpha_m \\
    & & & \beta_m
  \end{bmatrix}
  \
 = \left[  \begin{array}{cc} 
            T_{m} & \\
            \beta_{m} \cdot e_m^*  \end{array}
\right] \in \mathbb{R}^{(m+1) \times m} 
\]
with $T_{m}$ a (real) symmetric tridiagonal matrix. 

Let $g(t) = \sum_{i=1}^p \frac{\omega_i}{t-\sigma_i}$ be the Zolotarev approximation to $t^{-1/2}$. We get the $m$-th
Lanczos approximation to $g(Q^2)(Qb)$, which in turn approximates $\sign(Q)b$,  by running a multishift cg method, based on the Lanczos process, for the $p$ systems $(Q^2-\sigma_i I) x_i = Qb$. This is summarized as Algorithm~\ref{algo:cg-lanczos}, where $A = Q^2, c = Qb$. Herein, the factors $\rho_m^{(i)}$ are the scaling factors between the Lanczos vector and the residuals, see \cite{Paige.Parlett.vdVorst.95}: 
\begin{equation} \label{rho:eq}
r_m^{(i)} = Qb-Q^2x_m^{(i)} = \rho_m^{(i)} v_{m+1}, \enspace \mbox{ and } \rho_m^{(i)} = (-1)^m \|r_m^{(i)} \|_2.
\end{equation}

\begin{algorithm}
  \caption{Multishift cg}
  \label{algo:cg-lanczos}
  set $x_{-1} = 0$, $\rho_0^{(i)} = \| c \|_2 $, $ \tau_0^{(i)} = 1$, $v_1 = (1/\|c\|)c$ \;
  \For{ $j = 0, 1, \dots$ }{

    compute $\alpha_{j+1}$, $\beta_{j+1}$, $v_{j+2}$ using the Lanczos process for $A$
    \For{$i=1,\ldots,p$} {
       \If{ $j > 0 $ }{
          $\tau_j^{(i)} = \left[ 
           1 - 
            \frac{ \alpha_j - \sigma_i }{ \alpha_{j+1} -\sigma_i}
            \left(\frac{ \rho_j^{(i)} }{ \rho_{j-1}^{(i)} } \right)^2
            \frac{ 1 }{ \tau_{j-1}^{(i)} }
            \right]^{-1} $
        }
        $ \rho_{j+1}^{(i)} 
            = - \tau_j^{(i)} \rho_j^{(i)} \tfrac{ \beta_{j+1} }{ \alpha_{j+1} - \sigma_i}$\;
         $ x_{j+1}^{(i)} = \tau_j^{(i)} (x_j^{(i)} + \tfrac{1}{\alpha_{j+1}} r_j^{(i)}) + (1 - \tau_j^{(i)}) x_{j-1}^{(i)} $ \;
        $ r_{j+1}^{(i)} = \rho_{j+1}^{(i)} v_{j+2} $ \;
     }
    $x_m = \sum_{i=1}^p x_m^{(i)}$;
  }
\end{algorithm}

For the error $e_m$ of the $m$-th approximation $x_m$ we obtain 
\[  
   \underbrace{\sum_{i=1}^p \omega_i (Q^2-\sigma_i I)^{-1}(Qb)}_{:= x_*} -x_m =  \sum_{i=1}^p \omega_i (Q^2-\sigma_i I)^{-1}r_m^{(i)} 
    \, = \, \sum_{i=1}^p \rho_m^{(i)}\omega_i (Q^2-\sigma_i I)^{-1}v_{m+1},
\]
so we can express $\| e_m \|^2$ as 
\begin{equation} \label{error:eq}
\| e_m \|^2 = \|g_m(Q^2)v_{m+1}\|^2 = v_{m+1}^*g^2_m(Q^2)v_{m+1}, \mbox{ where } g_m(t) = \sum_{i=1}^p \frac{\rho_m^{(i)}\omega_i}{t-\sigma_i}.
\end{equation}
The elegant theory of moments and quadrature developed in \cite{golub-meurant93,golub-meurant97} allows to bound this quantity,
and more generally quantities of the form $v^*h(A)v$, from below and from above by performing some steps of the Lanczos process for $Q^2$ with starting vector $v_{m+1}$. The precise results is as follows:

\begin{theorem} \label{golub-meurant:thm}
Let $\hat{T_k}$ denote the tridiagonal matrix in the Lanczos relation \eqnref{lanczos_rel:eq} arising 
after $k$ steps of the Lanczos process with starting vector $v, \|v\| = 1$. Assume that $h: \mathR \to \mathR$ is at least 
$2k+2$ times continuously differentiable on an open set containing $[a,b]$, where $\spec(A) \subseteq [a,b]$.
\begin{itemize}
  \item[(i)] Approximating $v^*h(A)v$ with the Gauss quadrature rule using  $k$ nodes 
$t_j \in (a,b)$ gives
           \[
                  v^*h(A)v = e_1^*h(T^{\rm G}_k)e_1 + R^{\rm G}_k[h], \mbox{ where } T_k^{\rm G} = \hat{T}_k,
           \]
           with the error $R^{\rm G}_k[h]$ given as  
\begin{equation} \label{Gauss-err:eq}
   R^{\rm G}_k[h] = \frac{ h^{(2k)}(\xi)}{ (2k)! }
        \int_a^b 
          \left[ \prod_{j = 1}^k (t - t_j) \right]^2
          \ d \gamma (t),
          \quad a < \xi < b 
          \ .
\end{equation}
  \item[(ii)] Approximating $v^*h(A)v$ with the 
             Gauss-Radau quadrature rule using  $k-1$ nodes $t_j \in (a,b)$ 
             with one additional node fixed at $a$ gives 
             \[
                  v^*h(A)v = e_1^*h(T_k^{\rm GR})e_1 + R_k^{\rm GR}[h].
              \]
             Here, the tridiagonal matrix $T_k^{\rm GR}$ differs from $\hat{T}_k$ in that its $(k,k)$ entry $\alpha_k$ is 
             replaced by $\widetilde{\alpha}_k = a +\delta_{k-1}$, where $\delta_{k-1}$ is the last entry of the vector $\delta$ with $(\hat{T}_{k-1}-aI) \delta = \beta_ {k-1}^2 e_{k-1}$.
            The error $R^{\rm GR}_k[h]$ is given as  
\begin{equation} \label{Gauss-Radau-err:eq}
   R_k^{\rm GR}[h] = \frac{ h^{(2k-1)}(\xi)}{ (2k-1)! }
        \int_a^b 
          (t -a) \left[ \prod_{j = 1}^{k-1} (t - t_j) \right]^2
          \ d \gamma (t),
          \quad a < \xi < b 
          \ .
\end{equation}
\item[(iii)] Approximating $v^*h(A)v$ with the 
             Gauss-Lobatto quadrature rule using  $k-2$ nodes $t_j \in (a,b)$ and
             two additional nodes, one fixed at $a$ and one fixed at $b$, gives 
\[
                  v^*h(A)v = e_1^*h(T_k^{\rm GL})e_1 + R^{\rm GL}_k[h].
              \]
              Here, the tridiagonal matrix $T_k^{\rm GL}$ differs from $\hat{T}_k$ in its last column and row. With $\delta$ and $\mu$ the solutions of the system $(\hat{T}_{k-1}-aI)\delta = e_{k-1}$, $(\hat{T}_{k-1}-bI)\mu = e_{k-1}$ and $\widetilde{\alpha}_{k}, \widetilde{\beta}_{k-1}^2$ the solution of the linear system 
               \[
                 \left[ \begin{array}{cc} 1 & -\delta_k \\
                                          1 & -\mu_k
                        \end{array}
                  \right] \left[ \begin{array}{c} \widetilde{\alpha}_k \\ \widetilde{\beta}_{k-1}^2 \end{array} \right] = 
                   \left[ \begin{array}{c} a \\ b \end{array} \right] ,
               \]

               the tridiagonal matrix $T_k^{\rm GL}$ is obtained from $\hat{T}_k$ by replacing $\alpha_k$ by 
               $\widetilde{\alpha}_k$ and  $\beta_{k-1}$ by $\widetilde{\beta}_{k-1}$. The error $R^{\rm GL}_k[h]$ is given as  
\begin{equation} \label{Gauss-Lobatto-err:eq}
   R_k^{\rm GL}[h] = \frac{ h^{(2k-2)}(\xi)}{ (2k-2)! }
        \int_a^b 
          (t -a)(t-b) \left[ \prod_{j = 1}^{k-2} (t - t_j) \right]^2
          \ d \gamma (t),
          \quad a < \xi < b 
          \ .
\end{equation}
\end{itemize}
\end{theorem}

W apply Theorem~\ref{golub-meurant:thm} to the rational functions $h=g_m^2$ representing the error in \eqnref{error:eq}
Inspecting the terms $R^{\rm G}_k[h]$, $R^{\rm GR}_k[h]$ and $R^{\rm GL}_k[h]$ and noticing that $h^{(\ell)}(t) < 0 \; ( > 0)$ for $t \in [0,\infty)$ if $\ell$ is odd (even), we get the following corollary.

\begin{corollary} \label{bounds:cor}
In the case $h(t) = g_m(t)^2$ with $g_m$ from \eqnref{error:eq}, the estimates $e_1^*h(T_k^{\rm G})e_1$and $e_1^*h(T_k^{\rm GL})e_1$ 
from Theorem~\ref{golub-meurant:thm} (i), (iii) represent lower bounds, the estimate $e_1^*h(T_k^{\rm GR})e_1$ from (ii) 
represents an upper bound for the (square of the) error $\| x_m - x_* \|^2$. 
\end{corollary}

\section{Lanczos restart recovery}
To avoid ambiguities, let us call {\em primary} Lanczos process the one of the multishift cg method, i.e.\ the 
Lanczos process through which we obtain the approximations $x_m$.  The straightforward way to obtain the error 
estimates from Theorem~\ref{golub-meurant:thm} would be to perform $k$ steps of a new, {\em
restarted} Lanczos process which takes the current Lanczos vector
$v_{m+1}$ of the primary process as its starting vector. 
This results in the restarted Lanczos relation
\begin{equation} \label{eq:restartedlanczosrelation}
AV^{\rr}_{k} = V_{k+1}^{\rr}T_{k+1,k}^{\rr},
\end{equation} 
and we can now apply the theorem using the tridiagonal matrix
$T^{\rr}_k$ arising from the restarted process. This is,
however, far too costly in practice: computing the error estimate would
require $k$ multiplications with $A$---approximately the same amount of work
that we would need to advance the primary iteration from step $m$ to $m+k$.    

Fortunately, it is possible to cheaply retrieve the
matrix $T^{\rr}_k$ of the secondary Lanczos process from the matrix
$T_{m+1+k}$ of the primary Lanczos process. This {\em Lanczos restart recovery}
opens the way to efficiently obtain all the error estimates from
Theorem~\ref{golub-meurant:thm} in a retrospective manner: At iteration $m+k$
we get the estimates for the error at iteration $m$ without using any
matrix-vector multiplications with $A$ and with cost $\mathcal{O}(k^2)$, independently
of the system size $n$. 

For $m= 0,1,\ldots$, we define the tridiagonal matrix $T^{(m+1,k)}$ as 
the diagonal block of $T_{m+1+k}$ ranging from rows and columns $\max\{1,m+1-k\}$ to $m+1+k$. So
$T^{(m+1,k)}$ is a $(2k+1) \times (2k+1)$ matrix, except for $m+1 \leq k$, where its size is
$(m+1+k) \times (m+1+k)$.

The following theorem, see \cite{FrKaLiRi11}, shows that for Lanczos restart recovery we basically have to run the 
Lanczos process for the tridiagonal matrix $T^{(m+1,k)}$, starting with the $k+1$st unit vector $e_{k+1} \in \mathC^{2k+1}$. 

\begin{theorem} \label{lanczos_recovery:thm}
Let the Lanczos relation for $k$ steps of the Lanczos process for $T^{(m+1,k)}$ with starting vector 
$e_{k+1} \in \mathC^{2k+1}$ ($e_{m+1} \in \mathC^{m+1+k}$ if $m+1 \leq k$) be given
as
\begin{equation} \label{reduced-lanczos:eq}
  T^{(m+1,k)} \widetilde{V}_k = \widetilde{V}_{k+1,k} \widetilde{T}_{k+1,k}.
\end{equation}
Then the matrix $T_{k+1,k}^{\rr}$ of the restarted Lanczos relation \eqnref{eq:restartedlanczosrelation} 
is given as
\begin{equation} \label{recovery:eq}
T^{\rr}_{k+1,k} = \widetilde{T}_{k+1,k}.
\end{equation} 
\end{theorem}

The above theorem shows that we can retrieve $T_{k+1,k}^{\rr}$ from $T_{m+k+1,m+k}$ by performing $k$ steps of
the Lanczos process for the $(2k+1)\times(2k+1)$ tridiagonal matrix $T^{(m+1,k)}$. Herein, each step has work 
$\mathcal{O}(k)$, so that the overall cost for computing $T_{k+1,k}^{\rr}$ is $\mathcal{O}(k^2)$. So we conclude that the total cost for computing the error estimates from Theorem~\ref{golub-meurant:thm} 
is also $\mathcal{O}(k^2)$. 

\begin{algorithm}
  \caption{Lanczos approximation for Zolotarev function with error bounds }
  \label{LanczosPlusEstimates:alg}
  set $x_{-1} = 0$, $\rho_0 = \| b \|_2 $, $ \tau_0 = 1$ \;
  choose $k$ \;
  \For{ $m = 0, 1, \dots$ }{

    compute $\alpha_{m+1}$, $\beta_{m+1}$, $v_{m+2}$ using the Lanczos process for $A$ \;
    \For(\tcc*[f]{loop over poles}){ $i=1,\ldots,p$}{
    \If{ $m > 0 $ }{
      $\tau_m^{(i)} = \left[ 
        1 - 
        \frac{ \alpha_m -\sigma_i }{ \alpha_{m+1} -\sigma_i}
        \left( \frac{ \rho_m^{(i)} }{ \rho_{m-1}^{(i)} } \right)^2
        \frac{ 1 }{ \tau_{m-1}^{(i)} }
      \right]^{-1} $
    }
    $ \rho_{m+1}^{(i)} 
    = - \tau_m^{(i)} \rho_m^{(i)} \tfrac{ \beta_{m+1} }{ \alpha_{m+1} -\sigma_i}$\;
    $ x_{m+1}^{(i)} = \tau_m^{(i)} \left(x_m^{(i)} + \tfrac{\rho_m^{(i)}}{\alpha_{m+1}-\sigma_i} v_{m+1}\right) + \left(1 - \tau_j^{(i)}\right) x_{m-1}^{(i)} $ \;
   }
   $x_{m+1} = \sum_{i=1}^p \omega_i x_{m+1}^{(i)}$ \;
   \If{$m > k$} {
      perform $k$ steps of the Lanczos process for $T^{(m-k,k)}$ \;
      this yields the tridiagonal matrix $\hat{T}_k \in \mathC^{k \times k}$ \;
      $\ell_{m-k} = \| g_m(\hat{T}_k)e_1\|_2  $ \;
      $u_{m-k} =  \| g_m(\hat{T}^{\rm GR})e_1\|_2 $  \tcc*{$\hat{T}_k, \hat{T}^{\rm GR}$ given in Theorem~\ref{golub-meurant:thm}(ii)} \; 
   }
  }
\end{algorithm}

Algorithm~\ref{LanczosPlusEstimates:alg} shows how we suggest to use the results exposed so far. It computes the Lanczos
approximations $x_m$ for $g(A)b$ with $g(t) = \sum_{i=1}^p
\frac{\omega_i}{t-\sigma_i}$ and bounds $\ell_{m-k}, u_{m-k}$ for the error at
iteration $m$ based on the Gauss and the Gauss-Radau rule. The Algorithm can be modified to also obtain error estimates or bounds based on the Gauss-Lobatto rule and to get bounds for the $A$-norm in case we deal with a linear system.

\section{Numerical results}

In this section we report the results of several numerical experiments with relatively 
small lattices of size $8^4$ to $16^4$. In our computations we used the common deflation 
technique as described, e.g.\ in \cite{vdE02}: 
We precompute the first, $\lambda_1,\ldots,\lambda_q$ say, eigenpairs of smallest modulus. 
With $\Pi$ denoting the orthogonal projection onto the space spanned by the corresponding
eigenvectors, we then have $\sign(Q)b = \sign(Q(I-\Pi)b) + \sign(Q\Pi b)$. Herein, we know
$\sign(Q\Pi b)$ explicitly, so that we now just have to approximate $\sign(Q(I-\Pi)b)$.
In this manner, we effectively shrink the eigenvalue intervals for $Q$,
so that we need fewer poles for an accurate Zolotarev approximation and,
in addition, the linear systems to be solved converge more rapidly.
Within an iterative solver for the overlap operator this approach results in a major speedup,
since $\sign(Q)b$ must usually be computed repeatedly for various vectors $b$. 
For Algorithm~\ref{LanczosPlusEstimates:alg} it has the additional advantage that we immediately
have a very good value for $a$, the lower bound on the smallest eigenvalue of $Q^2$ for which we can take $\lambda_q^2$.
In all our computations we deflated the smallest 30 eigenvalues, and we chose the Zolotarev approximation
to have error less than $10^{-9}$. 

\begin{figure} 
\centerline{\includegraphics[width=0.45\textwidth,height=0.21\textheight]{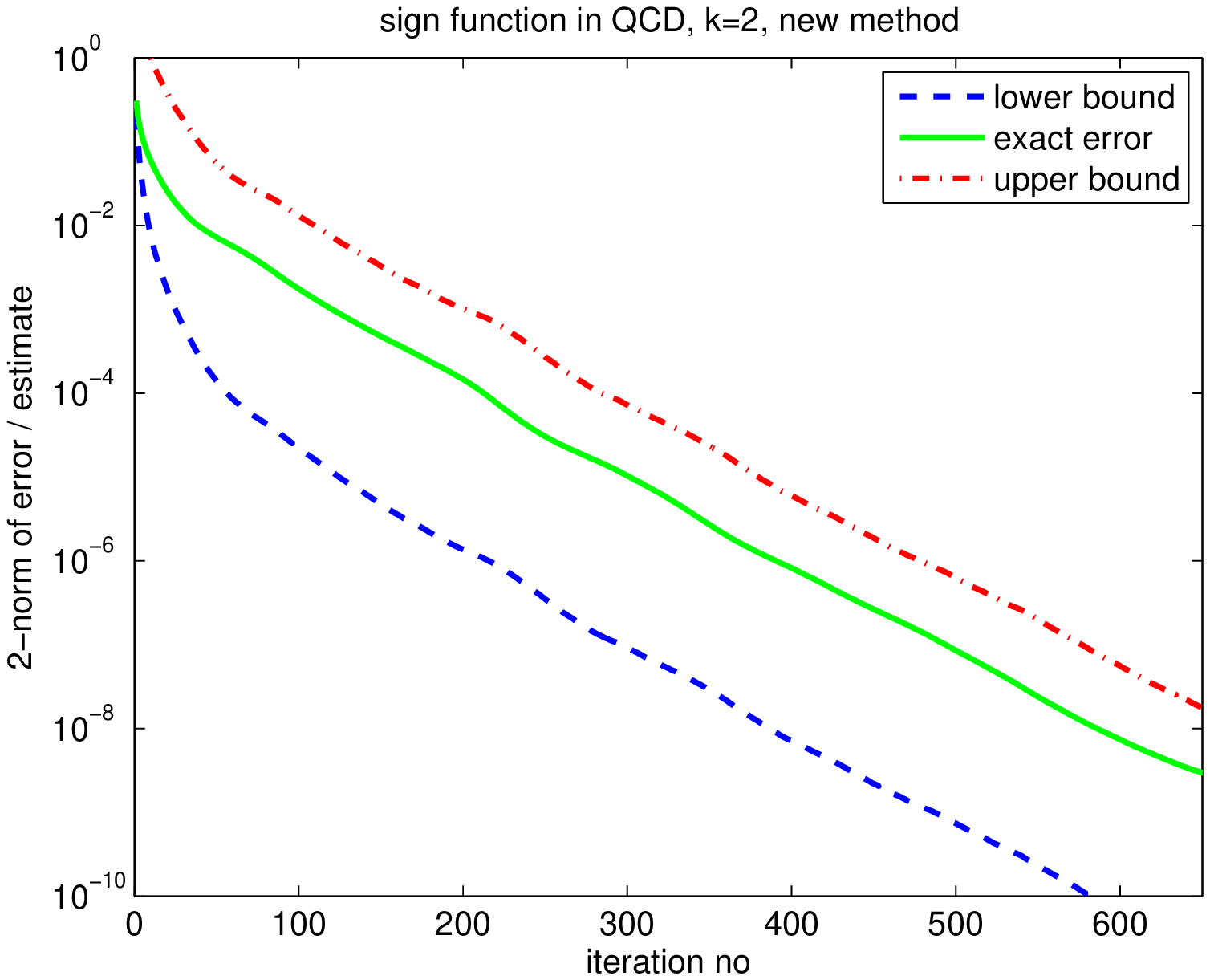}
\hfill \includegraphics[width=0.45\textwidth,height=0.21\textheight]{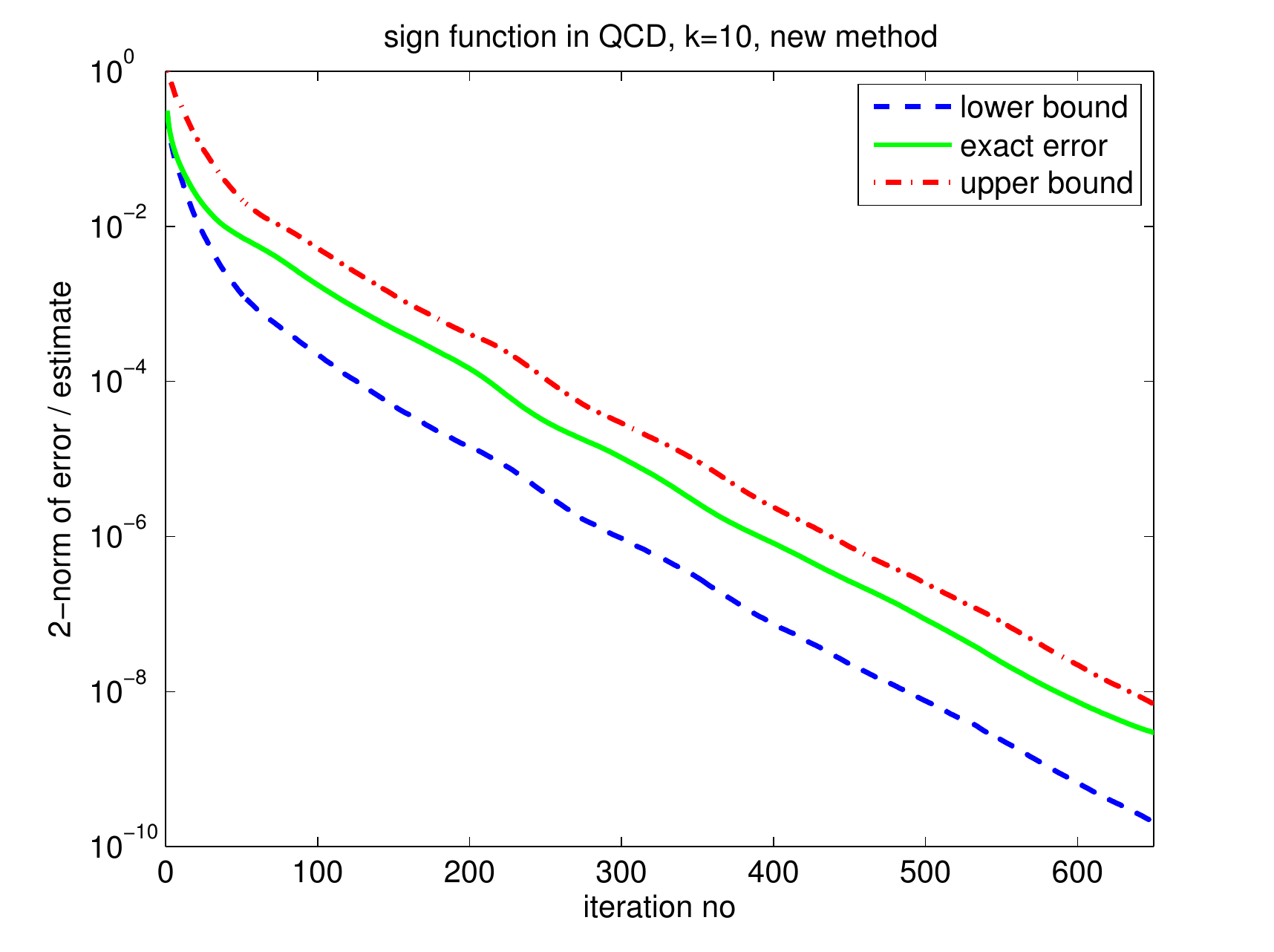}}
\centerline{\includegraphics[width=0.45\textwidth,height=0.21\textheight]{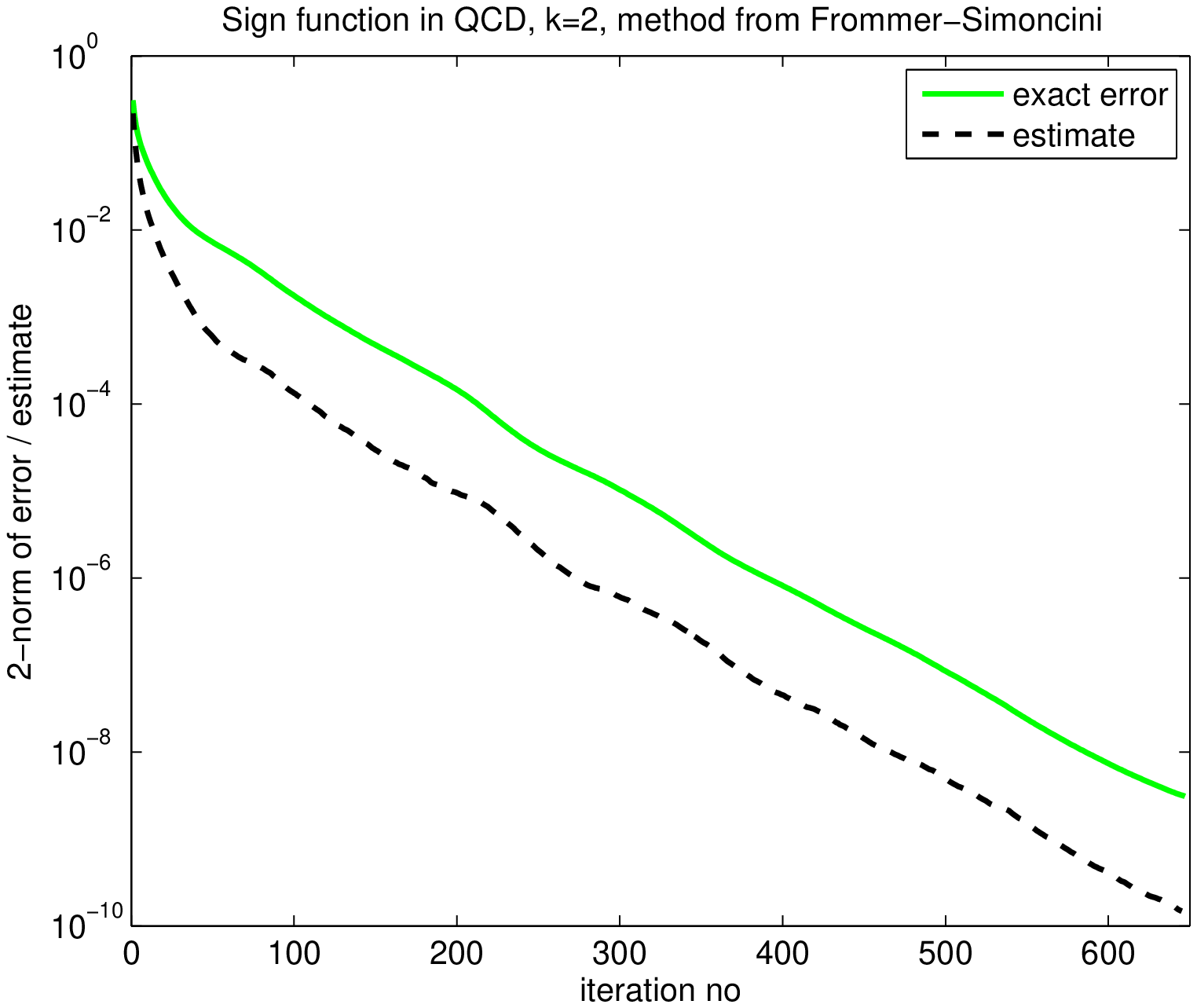}
\hfill \includegraphics[width=0.45\textwidth,height=0.21\textheight]{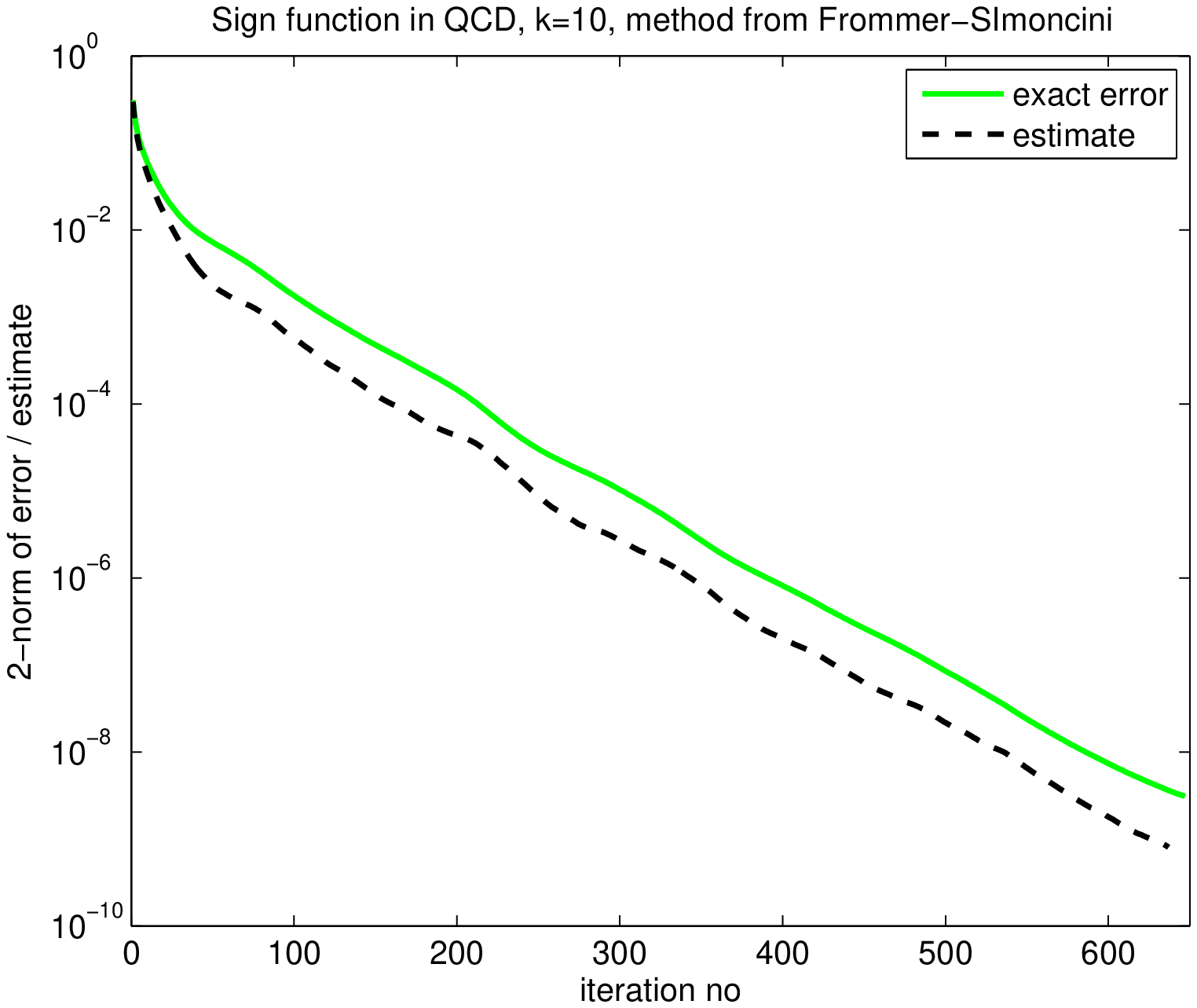}}
\caption{Error bounds and exact error
for Zolotarev approximation for $\sign(Q)$,
$8^4$ lattice. Left column: $k=2$, right column: $k=10$. Top row:
Algorithm~\protect\ref{LanczosPlusEstimates:alg}, bottom row: method from
\cite{FrSi07}.
}
\label{qcd_8:fig}
\end{figure}

Figure~\ref{qcd_8:fig} shows results for the $8^4$ configuration
available in the matrix group {\tt  QCD} at the UFL sparse matrix collection \cite{UFL} 
as matrix {\tt conf5.4-00l8x8-2000.mtx}. This is a dynamicyally generated configuration at $\beta=5.4$.
The (effective) condition number of the (deflated) matrix $Q^2$ is approximately $4.5\cdot 10^4$.
The left column of the figure reports upper and lower bounds from Algorithm~\ref{LanczosPlusEstimates:alg} whereas the right columns gives the estimates from earlier work \cite{FrSi07} which are known know to be lower bounds. 
The top row takes $k=2$ in Algorithm~\ref{LanczosPlusEstimates:alg} (and a similar parameter in the method from
\cite{FrSi07}), and the bottom row refers to $k=10$. We see that going from $k=2$ to $10$ results in a significant 
gain in accuracy and that for $k=10$ the upper and lower bounds just differ by a factor of 10. 
 
\begin{figure} 
\centerline{\includegraphics[width=0.45\textwidth,height=0.21\textheight]{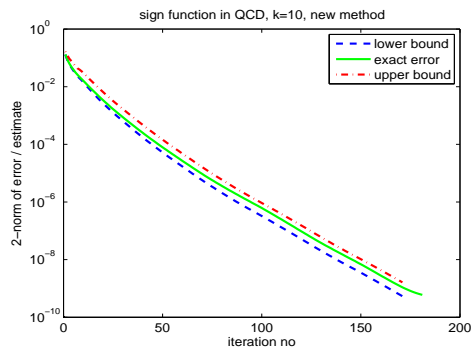}}
\caption{Error bounds and exact error for Zolotarev approximation for $\sign(Q)$, 
$16^4$ lattice, Algorithm~\protect\ref{LanczosPlusEstimates:alg}.}
\label{qcd_16:fig}
\end{figure}

Figure~\ref{qcd_16:fig} gives the results for Algorithm~\ref{LanczosPlusEstimates:alg} with $k=10$ for a configuration on a $16^4$ lattice. The configuration was the result of a quenched simulation.
The condition number of the deflated matrix $Q^2$ is now $64^2$, i.e.\ less than 
for the $8^4$ lattice. Therefore, the convergence speed as well as the quality of the bounds are better than 
for the $8^4$ lattice.  

\bibliographystyle{abbrv}
\bibliography{error_bounds}
\end{document}